I. Gódor, M. Luvisotto, S. Ruffini, K. Wang, D. Patel, J. Sachs, O. Dobrijevic, D. P. Venmani, O. L. Moult, J. Costa-Requena, A. Poutanen, C. Marshall, J. Farkas, "A Look Inside 5G Standards to Support Time Synchronization for Smart Manufacturing," in *IEEE Communications Standards Magazine*, vol. 4, no. 3, pp. 14-21, September 2020, doi: 10.1109/MCOMSTD.001.2000010.

# A look inside 5G standards to support time synchronization for smart manufacturing

I. Gódor, M. Luvisotto, S. Ruffini, K. Wang, D. Patel, J. Sachs, O. Dobrijevic, D. Venmani, O. Le Moult, J. Costa-Requena, A. Poutanen, C. Marshall and J. Farkas

*Abstract*— Connectivity has a major role in current transformation of smart manufacturing. 5G is foreseen as integral part of an end-to-end networking infrastructure supporting smart manufacturing operation. The integration of 5G with Time Sensitive Networking is seen as a holistic communication solution for smart factories. The next generation of industrialization consists of real-time processes, automation, fixed and cellular networks and high diversity of devices. The integration of fixed and cellular devices in a single infrastructure to enable end-to-end deterministic communications requires high accuracy of time synchronization. The transport technologies in fixed networks have been evolving towards Time-Sensitive Networking and Deterministic Networking to fulfil the requirements for industrial communications. The latest cellular specifications for 5G are targeting ultra-reliable and low-latency communications that enable Industrial Internet of Things. This paper describes the state of the art for integrating TSN with 5G networks based on the support given in 3GPP TS 23.501 Release 16 [11]. The requirements for time synchronization in factory automation and integration with 5G networks are presented, together with the most recent advancements in the standardization process.

## I. INTRODUCTION

THE next generation of industrialization integrates fixed and cellular networking technologies to interconnect devices that require accurate timing information to provide reliable and low latency deterministic communications. The work in fixed networks has been progressing in the IEEE 802.1 Time-Sensitive Networking (TSN) Task Group [1] and the IETF Deterministic Networking (DetNet) Working Group [2]. A core set of TSN specifications are already available and applicable to multiple verticals. The first three generations of cellular networks have been designed mainly for consumer communications and specifically for downlink data transfer, i.e., asymmetric data communication. The latest specifications for 5G are targeting ultra-reliable and low-latency communications (URLLC) that enable Industrial Internet of Things (IIoT).

The industrial automation infrastructure demands reliable transport as well as seamless connectivity of sensors, actuators, and controlling devices distributed in both cellular and fixed networks. To address the needs of next generation industrial networks, high accuracy time synchronization and symmetric transport between fixed and mobile devices is required.

This paper reviews the requirements and the state-of-the-art in terms of time synchronization for industrial automation in integrated 5G-TSN networks. This paper describes the TSN capabilities in fixed networks and explains in detail how to provide similar capabilities via cellular networks and devices based on the support given in 3GPP TS 23.501 Release 16 [11].

The paper is structured as follows. Chapter II provides an overview of use cases and requirements for time synchronization in factory automation. Chapter III introduces the solutions defined in TSN for synchronization. The state-of-the-art in 5G for tackling accurate time synchronization is presented in Chapter IV. Chapter V highlights the main gaps between state-of-the-art and requirements and discusses future works in this area, while Chapter VI concludes this work.

## II. REQUIREMENTS FOR TIME-SYNC IN FACTORY AUTOMATION

Industrial applications require not only robust and reliable data transport, but often accurate time information for synchronous communications as well. Over time several technologies have been applied, but Ethernet-based networks are the most widely adopted infrastructure in industrial practice today. Such a proliferation of Ethernet stems from the facts that it is standardized and is a de-facto transport for TCP/IP-based industrial wireline communication, which also facilitates integration of industrial applications with Information Technology (IT) services. The use of Ethernet in industrial automation has been realized via protocols like EtherCAT, Ethernet Powerlink and PROFINET.

Time or clock synchronization is a key requirement of many applications deployed over industrial networks. It aims to harmonize the time among independent, often physically distributed, clocks of network devices. Even if the clocks are initially set to the same time, they drift eventually and accumulate possibly significant time deviations. Many aspects of every-day operation of a network rely on its devices to have a common timescale, e.g., for fault management purposes. There are different industrial applications in the factory automation domain that depend on the clocks to be tightly synchronized.

For example, a system of Programmable Logic Controllers (PLCs) usually has a shared automation objective and needs to implement control actions in a specific sequence. Therefore, the PLCs need to have their clocks synchronized. In such a case, the required time deviation (with respect to a common time reference) provided up to 100 PLCs should be at most around one microsecond [3]. Another example of stringent time accuracy requirement relates to closed-loop motion control. Motion control governs movement and rotation of machine parts in real-time, by regularly transmitting, e.g., values of requested speed and position to actuators, while also having sensors reporting on the actual values for these parameters. Both the transmission to all involved actuators and the reporting from all sensors must be accomplished at a specific moment in time to enable precise motion. Building on simultaneous communication under few tens of microseconds, a motion control system requests that up to 100 actuators and sensors have guaranteed clock synchronicity in the order of one microsecond, or even below [3]. High accuracy demand is also put forth by factory deployments of mobile robots, which aid, e.g., transportation of materials or manufacturing of products. These tasks often include a tight cooperation between multiple mobile robot units, for which robot actions need to be strictly synchronized. A summary of time/clock synchronicity demands in industrial automation is given in [3] and in Table 1. These requirements were major input for 3GPP SA2 standardization work when developing time synchronization support provided by 5G [11]. The time synchronicity given in Table 1 is defined as the maximum allowed time offset within a synchronization domain between the master clock and any individual device clock.

Table 1. Time synchronicity requirements in typical industrial automation scenarios

| Clock synchronicity accuracy level | Number of devices in one communication group for clock synchronization | Clock synchronicity requirement | Service area | Scenarios |
|---|---|---|---|---|
| 1 | Up to 300 UEs | < 1 μs | ≤ 100 m x 100 m | Motion control Control-to-control communication for industrial controller |
| 2 | Up to 10 UEs | < 10 μs | ≤ 2500m² | High data rate video streaming |
| 3 | Up to 100 UEs | < 1 μs | < 20km² | Smart Grid: synchronicity between PMUs |

### III. SYNCHRONIZATION IN WIRELINE NETWORKS AND TSN

To achieve time synchronicity among devices in factory automation, a mechanism is required to deliver proper time from one device to another. This generally comprises the exchange of several messages to enable an estimation of the latency that the network introduces. A device that network nodes synchronize their clocks to, is usually referred as the master device, while others are called slave devices. Many realizations of a time synchronization mechanism can be found today, such as the Network Time Protocol (NTP) [4] , the Precision Time Protocol (PTP) [5], and the generalized PTP (gPTP) [6], which can be considered as specific PTP profile defined in IEEE 802.1AS specifications and requires multicast transport. While NTP is used for wide-area networks such as the Internet, PTP is used, e.g., for telecom networks, gPTP focuses on networks with a limited coverage area and provides better time precision. gPTP may be used on Ethernet-based, optical and microwave networks, and in, for example, WiFi networks. Due to their higher synchronization accuracy, PTP and gPTP are adopted both in mobile backhaul networks and in time-sensitive industrial networks.

*A. Time synchronization in mobile backhaul networks*

In mobile networks, standard development organizations such as 3GPP and ITU-T play a major part to study and develop synchronization solutions and architectures. 3GPP typically defines timing accuracy requirements and standardizes solutions for mobile air interface. ITU-T also develops solutions and architectures related to 3GPP requirements. Specifically, the ITU-T Recommendation G.8275.1 Precision Time Protocol Telecom Profile [7] specifies a PTP profile that can be used for synchronizing time in wireline mobile backhaul networks. Such mechanism can also help achieving synchronicity in mobile industrial networks, where the timing from the wireline backhaul could be transported and distributed to the industrial network.

*B. Time synchronization in TSN*

The gPTP mechanism, defined in the IEEE 802.1AS standard [6], is a building block of the TSN standardization efforts carried out by the IEEE 802.1 TSN Task Group (TG). gPTP is explained in [8] and the feature topic on TSN of the IEEE Communications Standards Magazine [9] provides in-depth tutorials of TSN.

The initial goal of the TSN TG was to build audio and video bridges within a LAN environment, thus allowing audio and video devices to align in time the streams coming from various sources. This initial use case expanded towards a general add-on to traditional Ethernet to offer bounded and low latency, low packet delay variation (jitter) and extremely low packet losses for many use cases, including (but not limited to) industrial automation and automotive in-vehicle communication. Nowadays, the TSN toolbox includes more than 20 standards. Recently, TSN profile specifications (e.g., IEC/IEEE 60802) are being developed for selected use cases, e.g., industrial automation [9].

Industrial automation devices in a TSN network align their activities in time through a shared clock according to the gPTP protocol, thus forming a gPTP domain. The shared clock can either be a universal clock (also known as "wall clock"), or a working clock based on an arbitrary timescale. One node in the gPTP domain acts as the source of the shared clock and it is denoted as GrandMaster (GM). Clock information is distributed from the GM to all the nodes in the domain with dedicated messages defined by the gPTP protocol. To provide an example of clock distribution, consider a typical gPTP domain illustrated in Fig. 1.

The domain includes TSN end stations, one of which is the GM, as well as TSN bridges. In each link between two nodes, the node closer to the GM is called Clock Master (CM) and the other node is called Clock Slave (CS). The gPTP protocol is implemented in a distributed way by each node in the network.

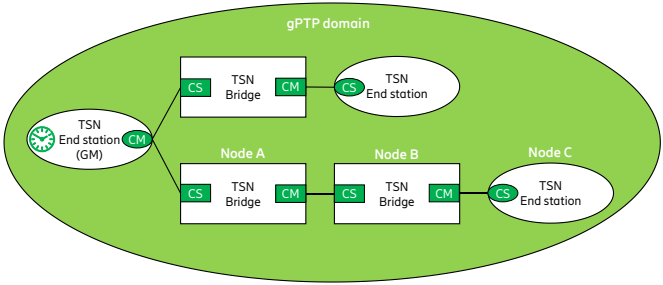

Fig. 1. Example of a typical gPTP domain where a clock is shared according to IEEE 802.1AS.

To provide an example, consider the following sequence of operations for Node B (TSN bridge) in Fig. 1:
1. Receive a gPTP Protocol Data Unit (PDU) from its direct CM (Node A) containing the time of the GM and the cumulative time delay from the GM.
2. Compute the propagation delay of the link with its direct CM.
3. Combine the information in 1 and 2 to compute the total delay from the GM and update its internal clock.
4. Compute its residence time, i.e., the time elapsed from the reception of the gPTP PDU (ingress) and its transmission (egress).
5. Send the gPTP PDU to its direct CSs (e.g., Node C), which contains the time delay from the GM (computed in step 3) and the residence time (computed in step 4).

IEC/IEEE 60802 includes time synchronization requirements for both working clock and universal clock. In the first case, the maximum deviation between a node's clock and the GM's working clock must be the range between 100 ns and 1 µs. In addition, redundant working clock domains can be setup for the same set of devices, so that zero failover time can be achieved in case a GM node experiences a failure. In practice, redundant domains can be realized through "cold standby" or "hot standby". In both cases, a node in the gPTP domain is ready to take the role as GM in case of failure of the primary GM. In the "hot standby" case, the back-up GM shares its clock through gPTP PDUs even when the primary GM is active. The current practice suggests the use of "hot standby" backup GMs for achieving the highest synchronization availability.

In case synchronization to a universal clock is implemented, the clock is provided by an external source (e.g., Global Positioning System) to the GM and shared within the plant via gPTP mechanisms. One possible use case for universal time synchronization is the need to record sequence of events by different devices within a plant, to enable root-cause analysis and plant optimization. In this use case, a more relaxed synchronization accuracy is allowed (up to 100 µs) and no redundant clock domain is required.

## IV. 5G TIME SYNCHRONIZATION SUPPORTING TSN

Integration of 5G with TSN to support time critical industrial applications requires end-to-end time synchronization, too. Time reference information is needed for the applications running in end devices in most of the industrial automation deployments. Time reference information is also required in the bridges of a TSN network when time-based TSN tools are used, like Scheduled Traffic (IEEE 802.1Qbv), to provide deterministic low latency for time-critical traffic.

### A. 5G internal system clock

Time synchronization is an integral part of the 5G cellular radio system and is essential for its operation. Time synchronization has been common practice already for cellular networks of different generations. The radio network components themselves are also time synchronized, e.g., for advanced radio transmission, such as synchronized Time Division (TDD) operation, cooperative multipoint (CoMP) transmission, or carrier aggregation. The new capability to provide reference time delivery as a service over the 5G system is designed to exploit the existing synchronized operation of the 5G radio access network. Such a building block enables end-to-end time synchronization for the communication of industrial applications running over 5G system.

The 5G internal system clock can be made available to a User Equipment (UE) with signaling of time information related to absolute timing of radio frames as described in 3GPP TS 38.331 [10]. 5G NR radio transmission is done in 10 ms radio frames, and each frame is identified with System Frame Number (SFN). These frames are used to define transmission cycles and they are also used to calculate accurate time reference over radio interface to a UE.

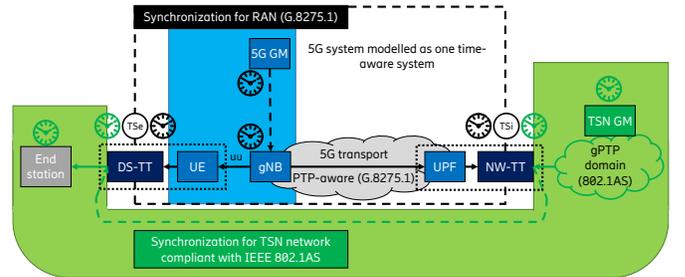

Fig. 2. Synchronization of TSN across 5G system.

The 5G system can use a 5G internal clock as time reference that is distributed over the transport network utilising PTP Telecom Profile [7] to its gNBs (5G base stations) as Fig. 2 shows. Based on the accurate time distributed over transport network, a gNB acquires the reference time value, further it modifies this value considering the propagation time to specific reference points in the system frame structure. Furthermore, a gNB transmits modified time reference value to UEs in either a System Broadcast (SIB) or a Radio Resource Control (RRC) message. It depends on the implementation what frequency is used to update the time reference at the end point connected to the UE, e.g., a TSN end station.

## B. 5G system enhancements to support gPTP

3GPP working groups conducted studies on how 5G can support the industrial automation vertical together with TSN including time synchronization. The 5G System (5GS) needs to interwork with the gPTP of the connected TSN network as gPTP is the default time synchronization solution for TSN-based industrial automation [12].

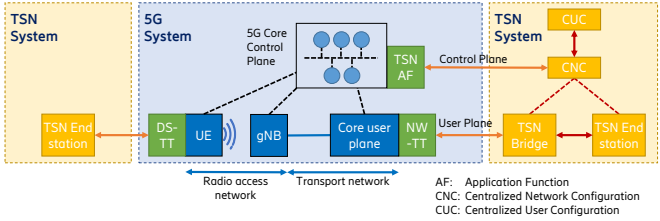

Fig. 3. 5G System as a bridge between two TSN systems.

The 5GS is considered as a virtual bridge for each User Plane Function (UPF) [11], which is connected to a TSN bridge. The UPF is a gateway that interconnects the 5G system to external networks, in this case a TSN network. The other end of the 5G virtual TSN bridge is at the UEs. Such a 5G virtual bridge is also considered as a time-aware system as per IEEE 802.1AS. Fig. 3 illustrates the new entities introduced at both edges of the 5G system to provide TSN Translation (TT) functionalities: Network side (NW-TT) at the UPF side and Device Side (DS-TT) at the UE side.

With the selected solution in the standard, the TSN timing information is delivered from NW-TT to DS-TT as gPTP messages via the 5G user plane. This approach is scalable, builds on the established 5G security model, and is flexible in various deployments, including network sharing and the support of non-public networks.

The 5G system is modelled as an IEEE 802.1AS time-aware system for supporting TSN time synchronization as specified in [11]. There are two synchronization processes running in parallel in an integrated 5G-TSN system: a 5G system synchronization process and a TSN synchronization process.
- *5GS synchronization*: the 5G GM provides the reference clock for 5GS internal synchronization, so the UPF and the UE are synchronized to the 5G reference clock.
- *TSN domain synchronization*: It provides synchronization service to devices in the TSN network. The gPTP Grand Master and the devices that are synchronized are outside the 5G system. Only the TSN translators at the edge of the 5G system track the gPTP domain. That is, only the TTs need to support IEEE 802.1AS operations, e.g., gPTP, timestamping and best master clock algorithm (BMCA).

The two synchronization processes can be considered independent from each other. For example, the gNB only needs to be synchronized to the 5G GM clock, the 5G synchronization process that guarantees RAN function working properly is kept intact and independent from the external gPTP synchronization process.

The independence of the two synchronization processes brings flexibility in time synchronization deployment, e.g., when upgrading an existing deployment. For example, if an operator has already deployed a 5G system in one area, in order to further support factories in the neighborhood with gPTP, only the UPF and UE side need additional enhancement and the overall 5G time synchronization remains untouched. Also, if 5G is added to a fixed TSN network with time synchronization, the TSN time synchronization is not modified due to the introduction of 5G.

## C. Downlink time synchronization

For every gPTP PDU entering the 5GS at the UPF, the NW-TT entity generates ingress Timestamps (TSi) based on the 5GS internal clock and embeds the timestamp in the gPTP message. The UPF forwards the gPTP message to the UE via the user plane (i.e., via a PDU session). UE receives the gPTP message and forwards it to the DS-TT. The DS-TT then creates an egress Timestamp (TSe) for the gPTP PDU of the external gPTP working domain. The difference between TSi and TSe is considered as the calculated residence time that the given gPTP message has spent within the 5GS. The DS-TT modifies the TSN timing information received from gPTP messages based on the calculated 5GS residence time and sends it towards to the next time-aware system.

## D. Support for multiple working clock domains

In smart manufacturing, multiple gPTP domains can coexist in a single shared communication network infrastructure where each gPTP domain consists of the commonly managed industrial automation devices which can have common timescale [9]. According to the ongoing standardization in [9] the minimum number of domains that a TSN industrial equipment needs to support depends on its device class. A constrained (currently called Class B) device has to support at least two gPTP domains: one working clock domain and one universal time domain. For redundancy, a feature rich (currently called Class A) device has to support at least two of each time domain, i.e., at least four gPTP domains altogether. 3GPP requires the 5G system in Release 16 to support networks with up to 32 working clock domains [3]. A gPTP working clock domain is characterized with a specific domain number in the gPTP PDUs. To support multiple gPTP domains in the 5GS, as shown in Fig. 4, the NW-TT generates an ingress timestamp (TSi) for every gPTP message carrying a specific domain number (where the domain number indicates a specific gPTP domain).

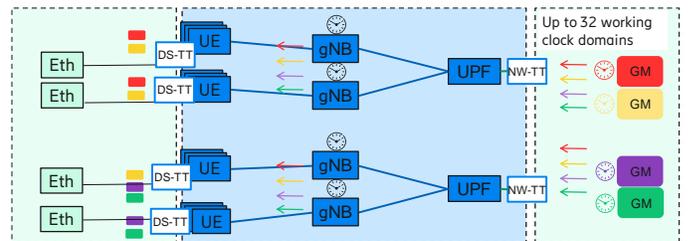

Fig. 4. Illustration of multiple working domains.

A UE receives gPTP messages and forwards them to the DS-TT. The DS-TT receives the original gPTP clock timing information and the corresponding TSi via gPTP messages for one or more working domains.

The DS-TT generates TSes for the gPTP PDUs for every external gPTP working domain. Ingress and egress time stamping are based on the 5G system clock to which the NW-TT and DS-TT are synchronized. An end station can select timing information of interest based on the domain number in the gPTP PDU.

A special case that is supported by the 5GS is when the 5GS clock acts as a general grandmaster and provides the time reference not only within the 5GS, but also to the rest of the devices in the deployment, including connected TSN bridges and end stations where, the time domains are merged into one common single time domain.

### E. Support for non-public networks and RAN sharing

3GPP defines Non-Public Networks (NPN) for industrial IoT in which the access to the NPN is restricted to a private group of authorized devices. There are two variants of NPN. A *standalone NPN* is a dedicated and isolated 5G network deployment for a private entity, like a factory. A *public network integrated NPN* is a private network that is partly sharing communication infrastructure with a public network.

It is also possible to share a common infrastructure for multiple NPNs. Normally, there are many companies in an industry park that provide common facilities to its enterprise customers. Each company can use a dedicated 5G NPN, where the core network nodes are deployed in the company premises and connected to, e.g., its local TSN infrastructure. In such case, a common 5G radio access network (RAN) can be used in the entire industry park in order to provide good coverage and capacity, meanwhile prohibit interference and coexistence challenges of multiple local RANs in a close vicinity. This RAN is shared among the multiple NPNs, but the NPNs are logically separated and capacity reservations can be provided per NPN. Each NPN can have its own working clock domains, which remain isolated, even if a shared RAN is used, as depicted in Fig 5. The proposed solution is not limited in the number of working domains that are supported, which means that each NPN can support multiple working domains independently without limitations of the 5G system.

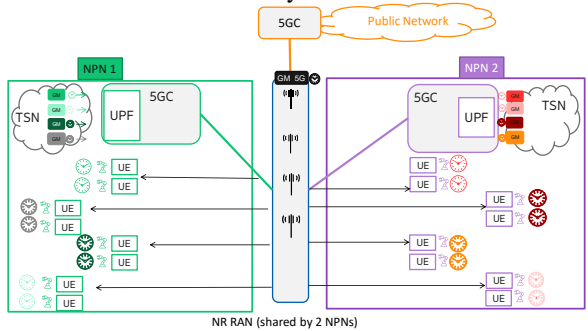

Fig. 5 Time synchronization in 2 NPNs with a shared RAN.

### F. Security aspects of time synchronization

In the 5G time synchronization, the gPTP messages are transmitted as user plane data. Therefore, all defined 5G security mechanisms that are used for data traffic also apply for the time synchronization data. Only authorized devices can participate in the 5G communication and all user data is encrypted. It is not possible for a device to monitor the time synchronization of a factory, even in a case like multiple NPNs sharing a common RAN, unless the device is a member of the NPN and included in the time synchronization communication.

What is available to any device within the coverage of the 5G RAN is the time used in the 5GS. However, this time does not reveal anything about the time that is used by any end device within any NPN. This functionality is commonly available in today's cellular networks, even if the timing is typically on much coarser level (for example, smart phones can receive the time-of-day from the network).

### G. Overhead for conveying gPTP messages

Time synchronization is typically applied in real-time industrial control applications like motion control. Let us consider a motion control application that has a user plane data rate of 50 Byte / 500 μs = 800 kbps. The gPTP data rate is around 2 * 50 Byte / 125 ms = 6.4 kbps. The overhead for 5GS to deliver the time sync message in user plane is 0.8%, and the overhead is constant regardless to the number of UEs. The 5GS is very well capable to handle a data rate of 6.4 kb/s for a UE, which traffic is marginal compared to the industrial URLLC traffic. In conclusion, the overhead of gPTP messages is insignificant to provide a time-synchronization service.

## V. GAP ANALYSIS AND FUTURE WORK

From the analysis of the state-of-the-art, a few areas emerge in which more work needs to be done. These areas, summarized in the following, will be the subject of future research and standardization efforts.

### A. Arbitrary placement of synchronization master

When bringing 5G into an industrial setting, it may be of interest that a timing GM can be connected wirelessly to the system. When a gPTP GM is at UE side, the 5G system needs to support the TSN timing distribution both in uplink direction (UE-to-UPF synchronization) and to other UEs (UE-to-UE synchronization). Although the support for TSN timing distribution in the uplink is only considered for Release 17, it is important to note that the above presented 5G time synchronization solution is well prepared for this functionality, because the gPTP working clock domain is distributed in the user-plane. The UE-to-UE synchronization case introduces more stringent accuracy requirement on the synchronization performance, since it doubles radio propagation delay over the air interface. Further enhancement of 5GS is required in order to fulfill the industrial automation synchronicity requirement for the UE-to-UE synchronization case. Furthermore, the downlink and uplink time synchronization are independent of each other via timestamping and determination of 5GS residence time.

### B. Synchronicity budget for 5G system

Synchronization for Industry Automation applications, such as motion control, allows maximum 1 μs synchronicity [3]. The requirement is between the sync master and any device of the clock domain, which is an end-to-end (E2E) requirement [13]. A TSN network can have multiple hops in the E2E chain. When 5GS is modelled as a virtual bridge interworking with other TSN nodes, the 5GS only contributes to a fraction of the 1 μs

E2E synchronicity budget, therefore, the actual synchronicity budget left for 5GS is even less. This can be partially compensated since the 5GS replaces multiple wired network hops with reliable wireless connectivity, simplifying the deployment in the factory. A reasonable synchronicity budget for 5GS in real industrial automation scenarios needs to be further investigated and activities are planned on 3GPP Rel-17 [14] in this domain.

*C. UE relative timing*

Once provided with absolute timing from the gNB as 5G reference time, multiple UE devices will be synchronized together, nominally to the same time.

In the 5G radio network the precision is determined by

1. The quantization of the absolute time information provided [10].

2. The difference in the propagation time between each of the UEs and their respective gNB antennas.

Uncertainty in the synchronization of the UEs is introduced by the difference in the delay resulting from the radio signal propagation to the UE, which depends on the relative locations of the UE devices and their respective gNB. If the UE devices happen to be the same distance from their gNB but in different directions, they can still experience the same propagation time of flight delay. Clearly if the UE devices are moving then their relative timing can change.

The differences in the propagation time are largely countered in 5G by the Timing Advance mechanism, a control loop which aligns the timing of the incidence of the uplink signals from the UE as they arrive at the gNB. One major contribution to the error in this mechanism is due to quantization, in 5G of 32 ns or more depending on the numerology of the actual radio configuration. This appears as a source of error in the synchronization between the UEs, whatever the relative separation of the UE devices [15].

The UE radio layer needs to be smarter to distinguish attenuation, packet loss, signal reflections and other propagation delays. Directional antennas with beam forming and other methods to assist the signal to find the best path can be explored for increased timing accuracy.

Taking these two factors into account the relative timing accuracy between two local UE devices connected to the same serving cell should still be within 1 µs.

## VI. CONCLUSION

This paper explains how 5G supports time synchronization for industrial automation even in integrated deployments with wireline networks. As future wireline industrial communication networks will be based on TSN, the paper focuses on the integration between 5G and TSN and discusses several aspects that need to be considered for accurate time synchronization between the two domains. If DetNet is also used, then the time synchronization provided by 5G and TSN will apply to DetNet as well because DetNet leverages time synchronization provided by the sub-network technology. We describe how the 5G system internal synchronization is maintained and leveraged, furthermore, what specific enhancements have been added to 5G to support the time synchronization protocol in the wireline domain, i.e., gPTP. Support for multiple clock domains, time synchronization in non-public 5G networks, and security aspects are also explained. The paper also discusses aspects that will be refined by 3GPP Release 17 work, such as the possibility of connecting a gPTP time source via 5G and potentially consider relative synchronization between different UEs.

ACKNOWLEDGMENT

Authors thank John Diachina, Magnus Sandgren, Mårten Wahlström, Anders Höglund, Torsten Dudda, Shabnam Sultana, and Zhenhua Zou, who contributed to the standardization of the 5G time synchronization concept described in this article.

Part of this work has been performed in the framework of the H2020 project 5G-SMART co-funded by the EU. Authors would like to acknowledge the contributions of their colleagues from 5G-SMART although the views expressed are those of the authors and do not necessarily represent the views of the 5G-SMART project.

REFERENCES

[1] IEEE 802.1 Time-Sensitive Networking (TSN), http://www.ieee802.org/1/tsn
[2] IETF Deterministic Networking (DetNet), https://datatracker.ietf.org/wg/detnet/about/
[3] 3GPP TS 22.104 Service requirements for cyber-physical control applications in vertical domains
[4] RFC 5905 Network Time Protocol Version 4: Protocol and Algorithms Specification https://tools.ietf.org/html/rfc5905
[5] 1588-2008 - IEEE Standard for a Precision Clock Synchronization Protocol for Networked Measurement and Control Systems
[6] IEEE Std 802.1AS-2020 - IEEE Standard for Local and Metropolitan Area Networks - Timing and Synchronization for Time-Sensitive Applications
[7] G.8275.1: Precision time protocol telecom profile for phase/time synchronization with full timing support from the network https://www.itu.int/rec/T-REC-G.8275.1/en
[8] K. Stanton, "Distributing Deterministic, Accurate Time for Tightly Coordinated Network and Software Applications: IEEE 802.1AS, the TSN profile of PTP," in IEEE Communications Standards Magazine, vol. 2, no. 2, pp. 34-40, June 2018
[9] J. Farkas, L. L. Bello and C. Gunther, "Time-Sensitive Networking Standards," in IEEE Communications Standards Magazine, vol. 2, no. 2, pp. 20-21, June 2018
[10] 3GPP TS 38.331 Radio Resource Control (RRC) protocol specification https://www.3gpp.org/ftp/Specs/archive/38_series/38.331/
[11] 3GPP TS 23.501 System architecture for the 5G System https://www.3gpp.org/ftp/Specs/archive/23_series/23.501/
[12] J. Farkas, B. Varga, G. Miklós, and J. Sachs. "5G-TSN integration meets networking requirements for industrial automation " Ericsson Technology Review, vol. 96, no 7, 2019
[13] M. Khoshnevisan, V. Joseph, P. Gupta, F. Meshkati, R. Prakash and P. Tinnakornsrisuphap, "5G Industrial Networks with CoMP for URLLC and Time Sensitive Network Architecture," in IEEE Journal on Selected Areas in Communications, vol. 37, no. 4, pp. 947-959, April 2019
[14] 3GPP TR 22.832 "Study on enhancements for cyber-physical control applications in vertical domains", version 17.1.0
[15] A Mahmood et al., "Time Synchronization in 5G Wireless Edge: Requirements and Solutions for Critical-MTC", June 2019 https://arxiv.org/pdf/1906.06380

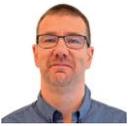
**Dr. István Gódor** is master researcher at Ericsson Research, a member of IEEE and Committee on Telecommunication Systems of the Hungarian Academy of Sciences (MTA).

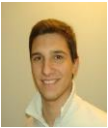
**Dr. Michele Luvisotto** is a scientist at ABB Corporate Research Center in Västerås, Sweden. He is leading research projects aimed at high-performance wireless communications for industrial applications.

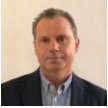
**Mr. Stefano Ruffini** is an Expert at Ericsson Research. He is the Rapporteur of the ITU-T SG15 Q13 and Chair of the ITSF (International Timing & Sync Forum).

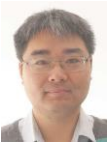
**Dr. Kun Wang** is a senior researcher at Ericsson. He is leading a standardization on the concept development of 5G-TSN integration for vertical industries, as part of 3GPP System Architecture (SA2).

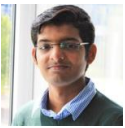
**Mr. Dhruvin Patel** is researcher at Ericsson where his current research is focused on 5G radio networks for factory automation. He holds an M.Sc. in System Engineering from Technical University of Ilmenau, Germany.

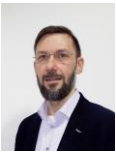
**Dr. Joachim Sachs** is a principal researcher at Ericsson Research. He coordinates the Ericsson 5G Industrial IoT research activities, standardization and industry collaborations.

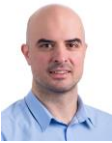
**Dr. Ognjen Dobrijevic** is a scientist at ABB Corporate Research in Vasteras, Sweden. He focuses on how to exploit 5G technologies for manufacturing and industrial automation.

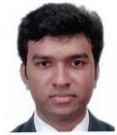
**Dr. Daniel Philip Venmani** is a senior research and standardization engineer at Orange Labs, Lannion. He represents Orange at ITU-T Study Group 15 and holds Ph.D. from UPMC-Paris VI.

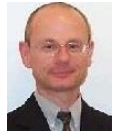
**Mr. Olivier Le Moult** is a senior metrology specialist at Orange Labs, Lannion. His works on emerging synchronization architectures, for example GNSS, optical and QKD architectures.

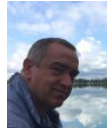
**Dr. Jose Costa-Requena** is Cumucore CEO and Staff Scientist at Aalto University. He worked at Nokia as Technology manager part of patent board and several years top inventor.

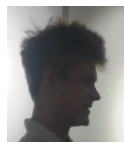
**Mr. Aapo Poutanen** is a senior engineer specialist in networking technologies at Cumucore. He holds M.Sc. in communications and networking from Aalto University, Finland.

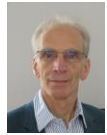
**Dr. Chris Marshall** is principal engineer at u-blox UK in Reigate, UK. He is leading and taking part in the development of cellular timing and positioning technologies.

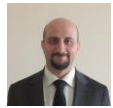
**Dr. János Farkas** is a principal researcher at Ericsson. He is the Chair of the IEEE 802.1 TSN TG and Co-Chair of the IETF DetNet WG. He holds Ph.D. in electrical engineering.